Title:

# On Strength of Brittle Nanomaterials: Confinement Effect on Weibull Distributions

*Dahye Shin[1] and Dongchan Jang[1]\**

*[1]Department of Nuclear and Quantum Engineering Korea Advanced Institute of Science and Technology (KAIST), 291 Daehak-ro, Yuseong-gu, Daejeon 34141, Republic of Korea*

e-mail:

- Dahye Shin : sth528@kaist.ac.kr

- **Dongchan Jang\* (corresponding author)** : dongchan.jang@kaist.ac.kr

**Abstract**

Recent progress in nanotechnology enables us to utilize the elastic strain engineering, the emerging technology capable of controlling the physio-chemical properties of materials via externally-imposed elastic strains, for hard materials. Because the range of accessible properties by imposing elastic strains are set by materials' elasticity limits, it is of great importance to suppress the occurrence of any inelastic deformations and failure, and thereby the fundamental knowledge on fracture behavior at nanoscale is highly required. The conventional Weibull theory, which has been widely used for last a few decades to explain the failure statistics of brittle bulk materials, has a limitation to be directly applied to the samples of nanometer dimensions because the baseline assumption on statistical equivalence becomes intractable for small samples. In this study, we suggest an integrated equation presenting the sample size effect on fracture strength for brittle nanomaterials by further considering the confinement of the flaw size distribution. This new approach is applicable to any homogeneous brittle nanomaterials whose failure is governed by linear elastic fracture mechanics, and shows good agreement with experimental data collected from literatures. We expect that this theoretical study offers new guideline to employ brittle nanomaterials in designing and fabricating the advanced strain engineering system.

**Keywords**

**Introduction**

Thermodynamic potentials and free energies of elastically-deformed solid bodies are known to explicitly depend on the strains or stresses [1,2]. This fact, in principle, offers a unique opportunity to make use of so-called strain engineering that enables materials engineers to tune and optimize many physical and chemical properties of functional materials by externally imposing elastic strains [3,4]. One required condition to take outmost benefits from the strain engineering is the large elastic deformability while suppressing the failure by inelastic deformation or fracture so as to maximize the accessible domains in materials' design parameter space. This condition is only marginally fulfilled for conventional hard materials, such as metals or ceramics, for their strain values at yield or fracture are typically less than 1% at most [5,6]. On the other hand, recent studies on nanomechanics revealed that both yield [7,8] and fracture [9,10] strengths of many hard nanomaterials drastically increase up to a significant fraction of their ideal strengths as the sample sizes decrease down to below a few hundred nanometers, and accordingly the elasticity limits increase as well. This strengthening effect is purely size-induced without involving any microstructural modification, and therefore gives us the additional possibility to broadly adopt the method of strain engineering for many hard materials as long as their dimensions manage to be at the nanoscale.

In most practical cases, fabrication of a specimen perfectly free from defects and flaws is almost unachievable, and therefore their strengths and elastic limits degrade substantially due to those imperfections [5]. In this regard, it is of great importance to properly understand and reliably predict occurrence of inelastic failure in the presence of pre-existing flaws in order to fully utilize the elastic strain engineering for hard materials. In general, under the scheme of linear elastic fracture mechanics (LEFM), Weibull analysis based on the weakest link theory well describes the strength and failure of brittle materials [11,12], in which the fracture strength scales with the sample volume following the inverse power-law relationship whose exponent is usually called Weibull modulus [12]. Its linkage to the weakest link theory is easily understandable once we consider the fact that the larger specimens likely contain more imperfections than the smaller ones, and hence have higher probability to include more fatal flaws, which is the largest according to LEFM [5]. However, not imparting the lower bound for the flaw sizes to be, this theory, in its original form, has some limitations to be directly used for very small samples in which the method of strain engineering is likely to work most efficiently. Conventional Weibull theory assumes sufficiently small flaws in comparison with the specimen to ensure the statistical equivalence of a randomly-chosen sub-sample, i.e. an arbitrary small volume defined as a part of the entire sample, but this condition becomes intractable for nanomaterials as their external dimensions decrease approaching the typical flaw sizes existing in the material. In this study, we derived an integrated equation describing the influence of extrinsic sizes on fracture strength of brittle nanomaterials, based on the conventional Weibull statistics but further considering the confinement effects on the flaw size distribution set up by the external dimensions. We confirmed the validity of this approach by comparing our predictions with the experimental data collected from literatures [13-15]. We expect that this theoretical study offers new guideline to design the enhanced strain engineered system composed of brittle nanomaterials.

**Extreme Value Distributions for Flaw Sizes and Fracture Strengths of Brittle Materials**

According to LEFM, the fracture strength of brittle materials, $\sigma_f$, scales with the reciprocal square root of the existing flaw size, $a$ [5] :

$$\sigma_f = \frac{K_{IC}}{\sqrt{\pi a}} \, F(\varphi). \tag{1}$$

, where $K_{IC}$ is the fracture toughness and $F(\varphi)$ is the correction factor considering the finite sample size effects and presented as a function of relative flaw size with respect to the sample dimension ($\varphi = a/t$, see the schematics in Fig 1). This LEFM formalism clearly indicates the explicit dependence of the fracture strength on the flaw size for a given $K_{IC}$, following the inverse square-root relationship, i.e., the larger the flaw is, the weaker the material becomes. In reality, a number of flaws with different sizes exist within the material, out of which the largest one determines the actual strength of the sample because the fracture initiates there at the lowest far-field load. In this sense, the strength of a brittle material is not an intrinsic property but rather stochastically determined by the statistical nature of flaw sizes. Because it is the largest flaw that matters for the fracture strength, but the smaller ones are of little interest, statistical model expressed in terms of the extreme values [12,16] needs to be taken into consideration in this study.

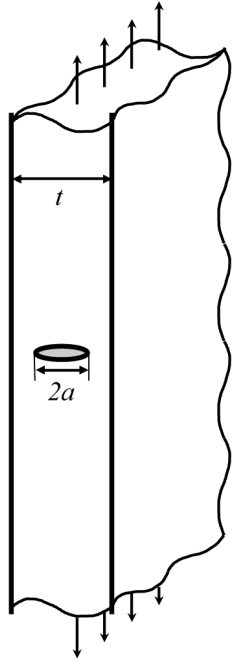

**FIG 1**. Schematic illustration of a cracked plate under uniaxial tensile load.

One of such formalisms, called Weibull statistics, is well recognized as an appropriate model for fracture strength of brittle materials due to its functional simplicity, physical satisfaction of zero lower bound, and most importantly excellent agreement with experimental data [12]. In his seminal work [11], Weibull established the extreme value formulation of fracture strength, in which the statistical random variable is parameterized by flaw strength, $\sigma_a$, the conceptual value associated with individual flaw by Eq. (1). Then, considering a brittle solid as being divided into many statistically-equivalent sub-samples of uniform volume, $V_0$ (see Fig 2(a)), the probability, $P_{V0}$, for this volume element containing $n$ discrete flaws not to fail under a given far-field stress of $\sigma$ is [17] :

$$P_{V0}(\sigma; \sigma_0, m) = \Pr\{\sigma_{\min} > \sigma\} = 1 - \Pr\{\sigma_{\min} \leq \sigma\} = \exp\left[-\left(\frac{\sigma}{\sigma_0}\right)^m\right]. \qquad (2)$$

, where Pr stands for the probability for the condition in the curly bracket to be true, $\sigma_{\min}$ is the minimum value out of all $n$ flaw strengths, $\sigma_a$, each of which is affiliated with the individual flaw present in the sub-sample of $V_0$, $m$ is the positive-valued parameter called shape parameter or Weibull modulus and $\sigma_0$ is the constant called scale parameter. Due to the one-to-one correspondence between the flaw strength and flaw size as in Eq. (1), the cumulative distribution function (CDF) in Eq. (2) can now be reformulated into the form having the maximum flaw size, $a_{\max}$, as the independent variable:

$$F_{V0}(a; a_0, m) = \Pr\{a_{\max} \leq a\} = \exp\left[-\left(\frac{a}{a_0}\right)^{-\frac{m}{2}}\right]. \tag{3}$$

, where $a_0$ is the scale parameter for flaw size distribution. Eq. (3) gives the probability of the largest flaw in $V_0$ to be smaller than a certain value $a$. As illustrated in Fig 2(a), the entire specimen with the finite volume $V$ can now be thought of as the juxtaposition of $N$ (=$V/V_0$) such sub-samples, all statistically equivalent, and its statistical characteristics can easily be built from that of the sub-sample using the max-stable nature of the extreme value distribution [16]. Namely, in order for the whole specimen to survive under a given far-field stress, all of the sub-samples should remain intact simultaneously, and therefore the probability for it to occur becomes [17] :

$$P_V(\sigma; \sigma_0, m) = P_{V0}(\sigma; \sigma_0, m)^N = \left\{\exp\left[-\left(\frac{\sigma}{\sigma_0}\right)^m\right]\right\}^{\frac{V}{V_0}} = \exp\left[-\frac{V}{V_0}\left(\frac{\sigma}{\sigma_0}\right)^m\right]. \tag{4}$$

The condition for two different specimens to have the same survival probability leads to the well-known Weibull scaling law:

$$\frac{\sigma_1}{\sigma_2} = \left(\frac{V_2}{V_1}\right)^{\frac{1}{m}}. \tag{5}$$

, where $\sigma_1$ and $\sigma_2$ are the strengths of each sample with volumes $V_1$ and $V_2$, respectively.

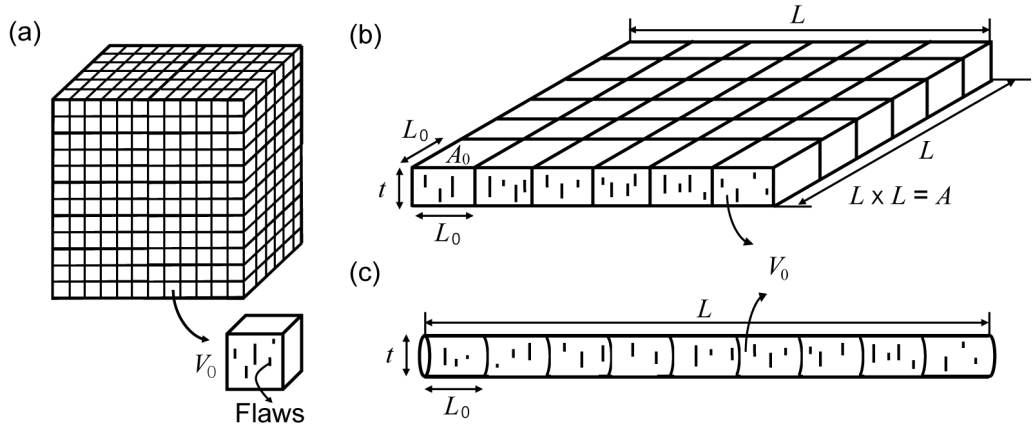

**FIG 2**. Schematic illustration for sub-sample division of (a) conventional bulk sample, (b) nano-plate, and (c) nano-wire into $V_0$.

**Confinement Effects on Extreme Value Distributions in Brittle Nano-samples**

From the perspective of weakest link theory, the conventional Weibull scaling law given in Eq. (5) emerges as a consequence of the fact that the larger specimen contains more flaws than the smaller one and therefore has higher probability for its weakest flaw, i.e. the largest, to be weaker than the one in the smaller. Here, one important condition to obtain such scaling relationship is the existence of the common reference sub-sample of $V_0$ ensuring the statistical equivalence across different specimens in a variety of external dimensions, because Weibull equation is built upon the statistical distribution formulated as the $N$-power of the CDF of a single sub-sample as in Eq. (4). Some requirements to keep such condition satisfactory may include i) materials synthesis in mutually comparable way between the different specimens so as for the fabrication processes not to affect the fundamental statistics and ii) more importantly, sufficiently small flaws compared with the whole specimen sizes so as to warrant the presence of the statistically-uniform sub-samples not influenced by the extrinsic dimensions of materials. However, the latter gets gradually intractable as the size of a specimen decreases approaching the typical flaw sizes, as is the case for nanomaterials. In that case, the universal reference sub-sample applicable to all specimens regardless of their external dimensions becomes hard to define. Instead, the coupling of fundamental statistics and the characteristic length, e.g., thickness of nano-plates or diameter of nano-wires as in Fig 2(b) and (c), needs to be taken into account, further requiring the modification of parameters for flaw size distributions, such as the one presented in Eq. (3).

The correlation between the flaw size distribution and characteristic length could be inferred by looking into two different statistical representations formulated with different random variables, i.e., one with the actual and the other with maximum flaw size in a sample. When sufficiently many flaws exist in a material, the probability for a flaw in a sub-sample of $V_0$ to have a certain size is usually given by a continuous probability density function (PDF), e.g. the Gaussian distribution. On the other hand, the size of the largest flaw determined in the sub-sample of $V_0$ can also serve as another random variable, i.e., the extreme value formulation, with which the statistical distribution follows the large-end tail of the parent distribution as shown in Fig 3(a) [18]. As long as the specimen is much larger than the width of the parent distribution, the statistical nature of the flaw size remains independent from the extrinsic dimensions of the material and the conventional Weibull scaling law still works well. However, provided that any flaw can never be larger than the whole body itself, the characteristic length of the sample must bound the width of the parent distribution as it becomes small to be comparable to the flaw sizes (Fig 3(b)), resulting in the mutual correlation of the former with the latter. Consequently, the extreme value distribution for the maximum flaw sizes should also be coupled with the characteristic length of the specimen, as schematically illustrated in Fig 3(b). In this study, in order to incorporate this coupling effect into statistical formulation, we assume that the scale factor $a_0$ in Eq (3) linearly scales with the characteristic length $t$, i.e., $a_0 = \alpha t$, where $\alpha$ is the proportionality constant. Replacing the scale parameter $a_0$ by the length dependent term, $\alpha t$, in Eq (3) and taking the derivative, the probability density function prescribing the largest flaw in the whole sample, $f_v(a,t;m)$, become:

$$f_V(a,t;m) = \frac{d}{da}\left\{\exp\left[-\frac{V}{V_0}\left(\frac{a}{\alpha t}\right)^{-\frac{m}{2}}\right]\right\} = \frac{m}{2\alpha t}\left(\frac{V}{V_0}\right)\left(\frac{a}{\alpha t}\right)^{-\left(\frac{m}{2}+1\right)}\exp\left[-\frac{V}{V_0}\left(\frac{a}{\alpha t}\right)^{-\frac{m}{2}}\right]. \quad (6)$$

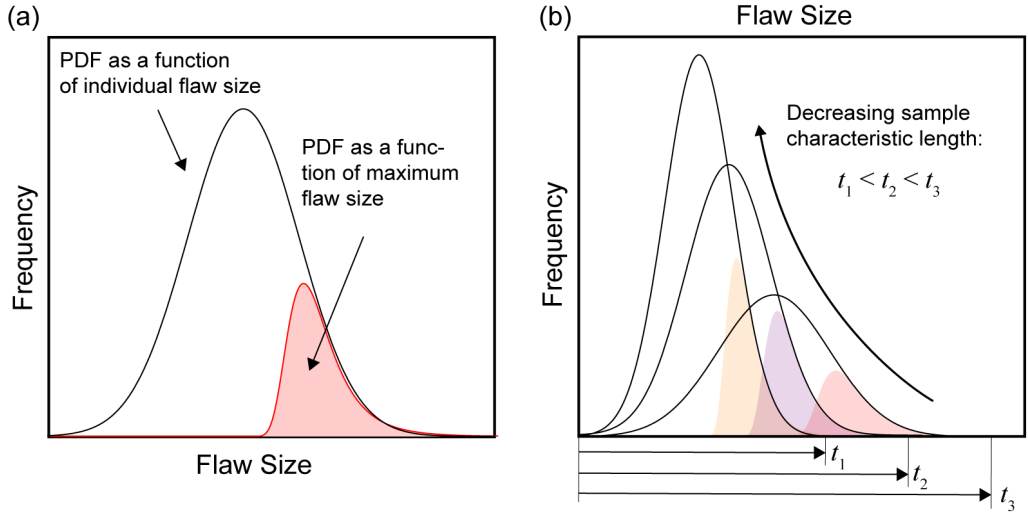

**FIG 3**. (a) Probability density distribution as a function of the maximum flaw size and its association with that of individual flaw size. (b) Schematic illustration showing the sample size confinement effects on the parent and maximum flaw size distributions.

The expectation value for the largest flaw size in $V$, $\bar{a}$, can be derived as follows:

$$\bar{a} = \int_0^\infty a\, f_V(a,t;m)\, da = \alpha t\left(\frac{V}{V_0}\right)^{\frac{2}{m}}\Gamma\left(1-\frac{2}{m}\right). \quad (7)$$

, where $\Gamma$ is the gamma function. Furthermore, plugging the maximum flaw size obtained in Eq. (7) into the LEFM strength equation given in Eq. (1), the expected value for the strength, $\bar{\sigma}$, of a brittle nanomaterial with volume $V$ is:

$$\bar{\sigma} = \frac{K_{\text{IC}}}{\sqrt{\pi\bar{a}}}F(\bar{\varphi}) = \frac{G(\beta,v)F(\bar{\varphi})}{\sqrt{\alpha\Gamma\left(1-\frac{2}{m}\right)}}\frac{K_{\text{IC}}}{\sqrt{\pi t}}\left(\frac{V_0}{V}\right)^{\frac{1}{m}}. \quad (8)$$

, where $\bar{\varphi} = \bar{a}/t = \alpha \, (V/V_0)^{2/m} \, \Gamma(1 - 2/m)$, and $G(\beta,\nu)$ is the geometric factor expressed as a function of the Poisson ration, $\nu$, and the inclination angle of the flaw with respect to the loading axis, $\beta$, which varies between 0 and $\pi/2$. When isotropically-distributed flaw orientation is assumed, $G(\beta,\nu)$ becomes $2/\pi$ with $\nu \sim 0.25$ [19,20]. Here, it is noteworthy that the inverse-square-root dependence of the fracture strength on the characteristic length of the specimen, $t$, newly appears as the result of confinement effect on the flaw size distribution in addition to the conventional Weibull scaling term, $(V_0/V)^{1/m}$. The approach presented in this study is applicable to any homogeneous brittle nanomaterials whose failure is governed by the LEFM-based Griffith fracture criterion [21]. The exact length scale below which the confinement effect on the Weibull distribution becomes non-negligible would vary, depending on the types of materials and quality of the synthesis process. In the following section, we confirmed the good agreement of our model with the experimental data collected from the samples under a few hundred nanometers [13-15].

**Case of 1D & 2D Nanomaterials**

The strength equations shown in Eq. (8) can be further developed for the 1D & 2D nanomaterials with simple geometry, such as nano-wires or nano-plates. With the help of illustrations given in Fig 2 (b) and (c), $V_0/V$ term that appears in Eq. (8) can be simplified into $L_0/L$ and $A_0/A$ for nano-wires and nano-plates, respectively, where $L$ and $L_0$ are lengths of whole and sub-sample of nano-wires, $A$ and $A_0$ are areas of whole and sub-samples of nano-plates, respectively. Then, according to Eq. (8) the mean strengths for each types of materials under the uniaxial tensile loading become:

$$
\begin{cases}
\bar{\sigma}_{1D}(t, L; m) = \dfrac{G(\beta,\nu)F(\bar{\varphi}(L;m))}{\sqrt{\alpha\Gamma\left(1 - \frac{2}{m}\right)}} \dfrac{K_{IC}}{\sqrt{\pi t}} \left(\dfrac{L_0}{L}\right)^{\frac{1}{m}}, \\[4mm]
\bar{\sigma}_{2D}(t, A; m) = \dfrac{G(\beta,\nu)F(\bar{\varphi}(A;m))}{\sqrt{\alpha\Gamma\left(1 - \frac{2}{m}\right)}} \dfrac{K_{IC}}{\sqrt{\pi t}} \left(\dfrac{A_0}{A}\right)^{\frac{1}{m}}.
\end{cases}
\tag{10}
$$

Furthermore, if we restrict our interest only to the set of samples whose axial lengths ($L$ for nano-wires) or areas ($A$ for nano-plates) remain constant, but only the characteristic lengths ($t$), i.e., diameter (nano-wires) or thickness (nano-plates), vary the mean strength becomes insensitive to the $L$ or $A$, but only depends on $t$, following the reciprocal square-root relation, i.e., $\sigma \propto t^{-1/2}$. This result is, in principle, equivalent with the scaling law offered by Gao, *et al*.

for brittle nano-plates [22]. In Fig 4, we present some examples of the actual experimental data collected from literatures, which report the uniaxial tensile strengths of brittle nano-whiskers made of ZnO [13,14] and single crystalline Cu [15]. Two important features are noteworthy here; i) clear demonstration of the reciprocal square-root relation between the diameter, $t$, and fracture strength, and ii) large scattering in the data. The former serves as the strong evidence for our work, and the latter indicates that the statistical fluctuation originated from the large Weibull modulus is still dominant in this regime.

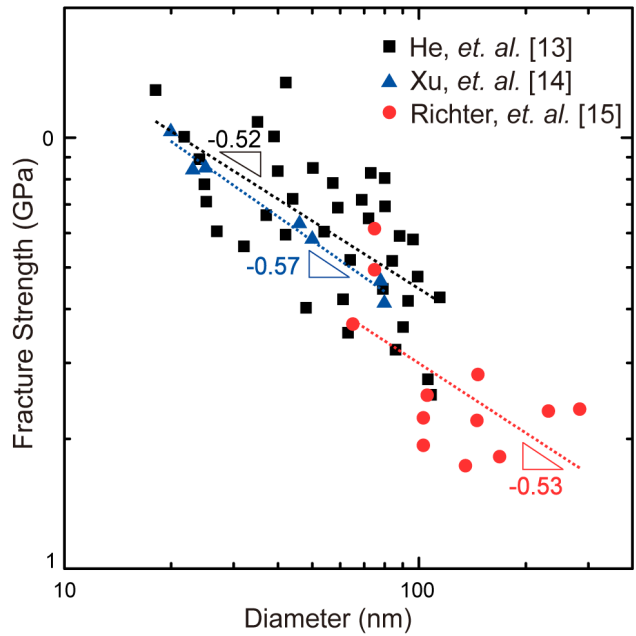

**FIG 4**. Fracture strength vs. diameter plots from tensile experiments on the brittle ZnO [13,14] and Cu [15] nanowires.

**Conclusions**

In conclusion, we suggest a new theoretical framework for fracture strength of brittle nanomaterials combining confinement effect on maximum flaw size distribution into the conventional Weibull statistics. By modifying the scale factor of the maximum crack size distribution function to be proportional to the characteristic length of the specimen, we successfully derive the integrated formula of fracture strength for nanomaterials which contains both the conventional Weibull and newly-added confinement terms in its scaling relationship. We further verified the validity of our equation in the range where the characteristic length is smaller than 300 nm by fitting it to actual experimental data collected for brittle nano-whiskers. This theoretical approach offers a foundation for the design of strain engineering and enables brittle nanomaterials to be more reliable and stable for their practical applications.

## Acknowledgements

The authors acknowledge financial support from National Research Foundation of Korea (NRF- 2014R1A2A2A01005009). All authors contributed to conceiving the research, establishing theoretical frameworks, analyzing the data, and writing the manuscript.